\documentclass[aps,prb,twocolumn,showpacs,citeautoscript,superscriptaddress]{revtex4-2}

\usepackage{amsmath,amssymb}
\usepackage{bm}
\usepackage{graphicx}
\usepackage{epstopdf} 
\usepackage{latexsym} 
\usepackage{subfigure}
\usepackage{color}
\usepackage{dsfont} 
\usepackage{wasysym}

\usepackage[colorlinks=true,linkcolor =black,citecolor=blue,urlcolor=blue,anchorcolor =blue]{hyperref}

\begin{document}

\title{Exchange renormalized crystal field excitation in a quantum Ising magnet KTmSe$_2$}

\author{Shiyi Zheng} 
\thanks{These authors contributed equally to this work}
\affiliation{State Key Laboratory of Surface Physics and Department of Physics, Fudan University, Shanghai 200433, China}
\affiliation{Shanghai Qi Zhi Institute, Shanghai 200232, China}
\author{Hongliang Wo} 
\thanks{These authors contributed equally to this work}
\affiliation{State Key Laboratory of Surface Physics and Department of Physics, Fudan University, Shanghai 200433, China}
\affiliation{Shanghai Qi Zhi Institute, Shanghai 200232, China}
\author{Yiqing Gu} 
\thanks{These authors contributed equally to this work}
\affiliation{State Key Laboratory of Surface Physics and Department of Physics, Fudan University, Shanghai 200433, China}
\affiliation{Shanghai Qi Zhi Institute, Shanghai 200232, China}
\author{Rui Leonard Luo} 
\affiliation{Department of Physics and HKU-UCAS Joint Institute for Theoretical and Computational Physics at Hong Kong, The University of Hong Kong, Hong Kong, China}
\author{Yimeng Gu} 
\affiliation{State Key Laboratory of Surface Physics and Department of Physics, Fudan University, Shanghai 200433, China}
\affiliation{Shanghai Qi Zhi Institute, Shanghai 200232, China}
\author{Yinghao Zhu} 
\affiliation{State Key Laboratory of Surface Physics and Department of Physics, Fudan University, Shanghai 200433, China}
\author{Paul Steffens} 
\affiliation{Institut Laue-Langevin, 71 Avenue des Martyrs, 38042 Grenoble Cedex 9, France}
\author{Martin Boehm} 
\affiliation{Institut Laue-Langevin, 71 Avenue des Martyrs, 38042 Grenoble Cedex 9, France}
\author{Qisi Wang} 
\affiliation{State Key Laboratory of Surface Physics and Department of Physics, Fudan University, Shanghai 200433, China}
\affiliation{Department of Physics, The Chinese University of Hong Kong, Shatin, Hong Kong, China}
\author{Gang Chen} 
\email{gangchen@hku.hk}
\affiliation{Department of Physics and HKU-UCAS Joint Institute for Theoretical and Computational Physics at Hong Kong, The University of Hong Kong, Hong Kong, China}
\author{Jun Zhao} 
\email{zhaoj@fudan.edu.cn}
\affiliation{State Key Laboratory of Surface Physics and Department of Physics, Fudan University, Shanghai 200433, China}
\affiliation{Shanghai Qi Zhi Institute, Shanghai 200232, China}
\affiliation{Institute of Nanoelectronics and Quantum Computing, Fudan University, Shanghai 200433, China}
\affiliation{Shanghai Research Center for Quantum Sciences, Shanghai 201315, China}

\date{\today}

\begin{abstract}
Rare-earth delafossite compounds, ARCh$_2$ (A = alkali or monovalent ion, R = rare earth, Ch = chalcogen), have been proposed for a range of novel quantum phenomena. Particularly, the Tm series, ATmCh$_2$, featuring Tm ions on a triangular lattice, serves as a representative group of compounds to illustrate the interplay and competition between spin-orbit coupling, crystal fields, and exchange couplings in the presence of geometric frustration. Here we report the thermodynamic and inelastic neutron scattering
studies on the newly discovered triangular-lattice magnet KTmSe$_2$.
Both heat capacity and neutron diffraction reveal the absence of long-range magnetic order.
Magnetic susceptibility shows strong Ising-like interactions with antiferromagnetic correlations.
Furthermore, inelastic neutron scattering measurements reveal a branch of dispersive crystal field excitations.
To analyze these observations, we employ both the transverse field Ising model and the full crystal field scheme, along with exchange interactions. Our results suggest a strong competition between spin exchange interactions
and crystal field effects. This work is expected to offer a valuable framework for understanding low-temperature magnetism in KTmSe$_2$ and similar materials.
\end{abstract}

\maketitle

\section{\label{sec:level1}Introduction}

Recent theoretical and experimental studies on the triangular-lattice magnets with
strong frustration and quantum fluctuation have proposed several possible realizations of
exotic quantum states~\cite{Balents2010,ZhouYi2017}.
Two common scenarios, involving triangular lattices occupied by either Kramers or non-Kramers ions, have been extensively investigated to elucidate unconventional magnetic ground states.
A well known example of the former case is the quantum spin liquid (QSL) material
YbMgGaO$_4$~\cite{LiYS2015,Shen2016,LiYS2016,LiYD2016-1,LiYD2016-2,Paddison2017,
Shen2018,LiYD2017-1,LiYD2017-2},
while TmMgGaO$_4$ with the integer spin local moments could be approximately
viewed as an example of the non-Kramers doublets~\cite{Liu2018,Shen2019,LiYS2020,Dun2021,Qin2022}.
Despite the successful introduction and description of the transverse field Ising model
(TFIM) in TmMgGaO$_4$,
further examination and research are necessary to address fundamental questions regarding the nature of magnetic excitations in generic non-Kramers triangular-lattice magnets~\cite{Shen2019,Liu2020,LiYS2020,Chen2019}.

Since crystal electric field (CEF) interactions play a vital role in determining both the magnetic ground state and single-ion anisotropy of rare-earth ions~\cite{Rau2018,Liu2019,Voleti2023}, it is essential to uncover the CEF spectra in order to understand the exotic quantum states found in triangular-lattice magnets. In addition to determining the CEF parameters and eigenvalues,
measuring the dispersion or splitting of the low-lying CEF states may
also provide useful information of low-temperature magnetism,
including dipole-dipole interactions~\cite{Wu2019},
phonon modes~\cite{Lummen2008,Maczka2008,Ruminy2016} and quadrupolar excitons~\cite{Iwasa2019}.

\begin{figure*}
\includegraphics[width=1\textwidth]{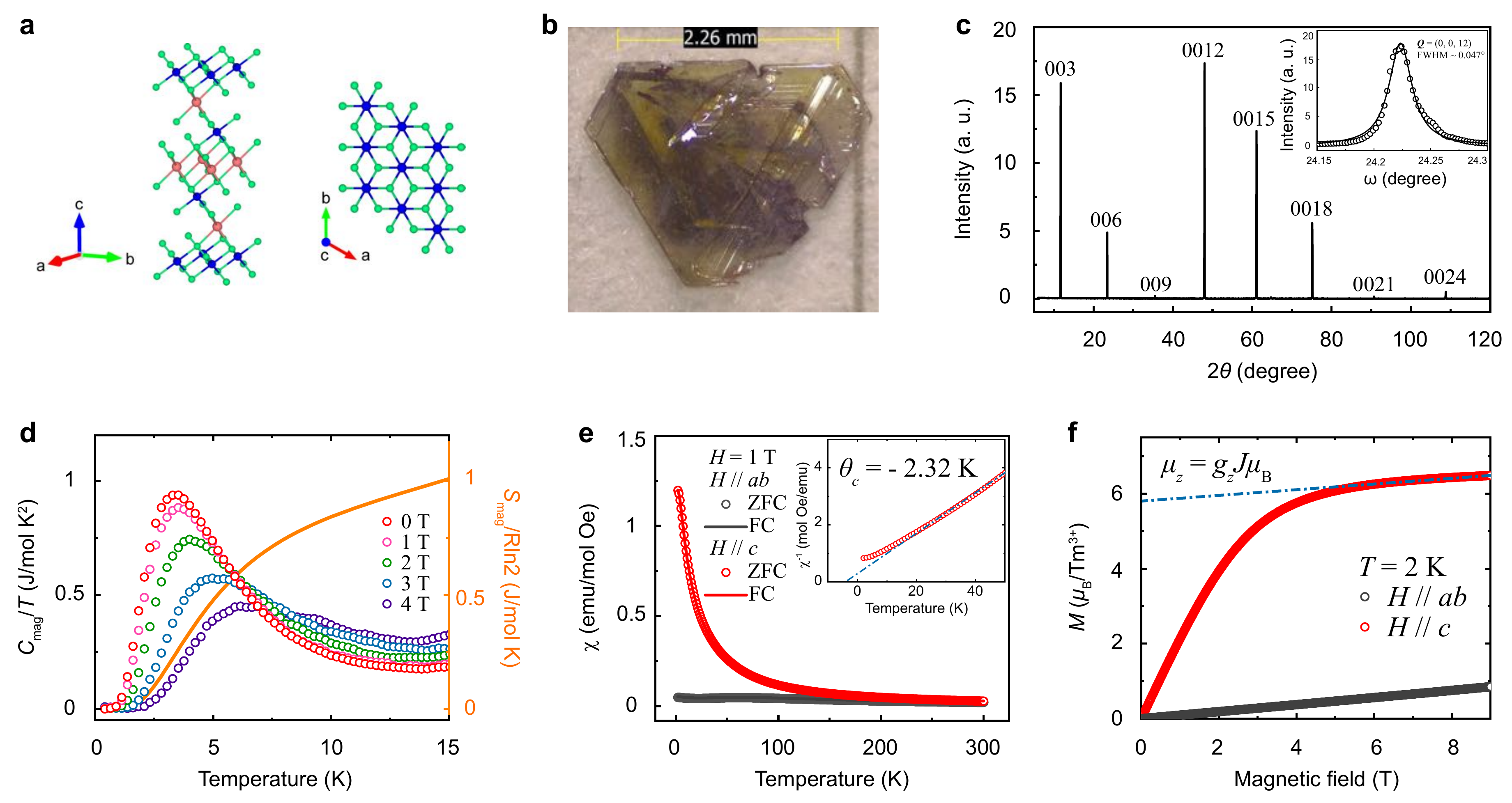}
\caption{
(a) Schematic of the crystal structure of KTmSe$_2$. Blue: Tm$^{3+}$, red: K$^{+}$, green: Se$^{2-}$. (b) Image of a typical single crystal. (c) X-ray diffraction pattern of KTmSe$_2$ single crystal on the (0, 0, $L$) plane, a.u. denotes arbitrary unit. (d) Heat capacity and entropy release of KTmSe$_2$ under different magnetic fields along the $c$ axis. (e) Temperature dependence of magnetic susceptibility from 2 K to 300 K for KTmSe$_2$ single crystal both in the $ab$ plane and along the $c$ axis under 1 T. The inset shows inversed magnetic susceptibility with negative Curie-Weiss temperature. (f) Isothermal magnetization of KTmSe$_2$ up to 9 T both in the $ab$ plane and along the $c$ axis at 2 K.
}
\label{fig:1}
\end{figure*}

In the last few years, a new class of triangular-lattice antiferromagnet ARCh$_2$ (A = alkali or monovalent ion,
R = rare earth, Ch = chalcogen) has attracted great attentions due to its potential in realizing novel quantum states~\cite{LiuWW2018}.
This family of materials crystallizes into a delafossite structure with space group $R$-$3m$ or $P6_3$/$mmc$,
while the trivalent R$^{3+}$ ions reside in a local environment with $D_{3d}$ symmetry.
Tremendous investigations on Yb$^{3+}$-~\cite{Bordelon2019,Ranjith2019-1,
Ding2019,Sarkar2019,Ranjith2019-2,Zhang2021-1,Dai2021,Scheie2023,Scheie2022,Xing2019-1},
Er$^{3+}$-~\cite{Xing2019-2,Xing2021,Ding2023}, and Ce$^{3+}$-based~\cite{Bastien2020,Kulbakov2021,Avdoshenko2022}
112-family magnets have been reported, including massive precise knowledge of the CEF excitations
in them~\cite{Bordelon2020,Baenitz2018,Zhang2021-2,Pocs2021,Gao2020,Scheie2020,Bordelon2021,Ortiz2022}.
However, while the majority of studies focus on materials in which the triangular lattice is formed by Kramers ions in ARCh$_2$ family, research on compounds featuring non-Kramers ions is rather limited. Unraveling the magnetic properties of these compounds remains an important task.
KTmSe$_2$, the new thulium-based 112-family material,
shares the same crystal structure as the QSL materials NaYbO$_2$~\cite{Bordelon2019} and NaYbSe$_2$~\cite{Dai2021}.
The perfect triangular-lattice layers of magnetic Tm$^{3+}$ ions are separated by the nonmagnetic K$^+$ layers,
resulting in a quasi-two-dimensional structure with geometrical frustration [see Fig.~\ref{fig:1}(a)].
The two lowest CEF levels of the non-Kramers Tm$^{3+}$ ion are point-group-symmetry-demanded singlets.
Nevertheless, the first excited CEF state is highly dispersive in the reciprocal space, and this is attributed
to the strong renormalization of the CEF excitation by the exchange interactions of the local moments.
The quantitative measurement of these excitations
provides a great chance to determine the exchange parameters
and the CEF energy levels.

In this paper, we report the thermodynamic, neutron scattering measurements and point-charge
calculations on the single-crystalline triangular-lattice magnet KTmSe$_2$ to investigate
the competion between the exchange coupling and the CEF effect by
directly revealing the exchange renormalized CEF excitations.
Heat capacity and neutron diffraction find the absence of long-range magnetic order in KTmSe$_2$
down to an ultra-low temperature of 60 mK.
The temperature dependent susceptibility and the field dependent magnetization show the Ising-like single-ion
anisotropy and predominantly antiferromagnetic spin correlations accompanied by a negative Curie-Weiss (CW) temperature.
Inelastic neutron scattering (INS) illustrates a branch of dispersive excitation from 0.85 to 1.6 meV
which is the first excited CEF state of Tm$^{3+}$,
indicating that a single-ion model can no longer describe the low-lying CEF excitations.
A frequently used solution is to apply the transverse field Ising model (TFIM) on KTmSe$_2$
to explain the dispersive magnetic excitation by constructing a pseudospin $S_{\text{eff}}$
= 1/2 from the ground state quasi-doublet~\cite{Liu2018,Shen2019,Qin2022,Chen2019,Liu2020}.
The effective model successfully explains the dispersive magnetic excitation
by introducing a transverse field perpendicular to the $z$ axis, and provides
a phase diagram that can include many different materials with similar Ising characters.
From a more complete perspective, establishing an effective magnetic Hamiltonian
consisting of all the CEF states and the spin exchange parameters simultaneously is also productive
and illustrative. As a complementary to the TFIM,
we then apply point-charge (PC) analysis and mean field-random phase approximation (MF-RPA)
to calculate the CEF spectra and low-energy magnetic excitations.

The remaining parts of the paper is organized as follows.
In Sec.~\ref{sec2}, we introduce the experimental details.
In Sec.~\ref{sec3}, we propose a TFIM for the low-lying doublets with the effective spin $S_{\text{eff}}$
to account for the dispersive CEF excitation.
In Sec.~\ref{sec4}, we include all the CEF states and the superexchange with the local $J$ moments to further
analyze the CEF excitations.
In Sec.~\ref{sec5}, we conclude with a discussion.

\begin{figure*}
\includegraphics[width=1\textwidth]{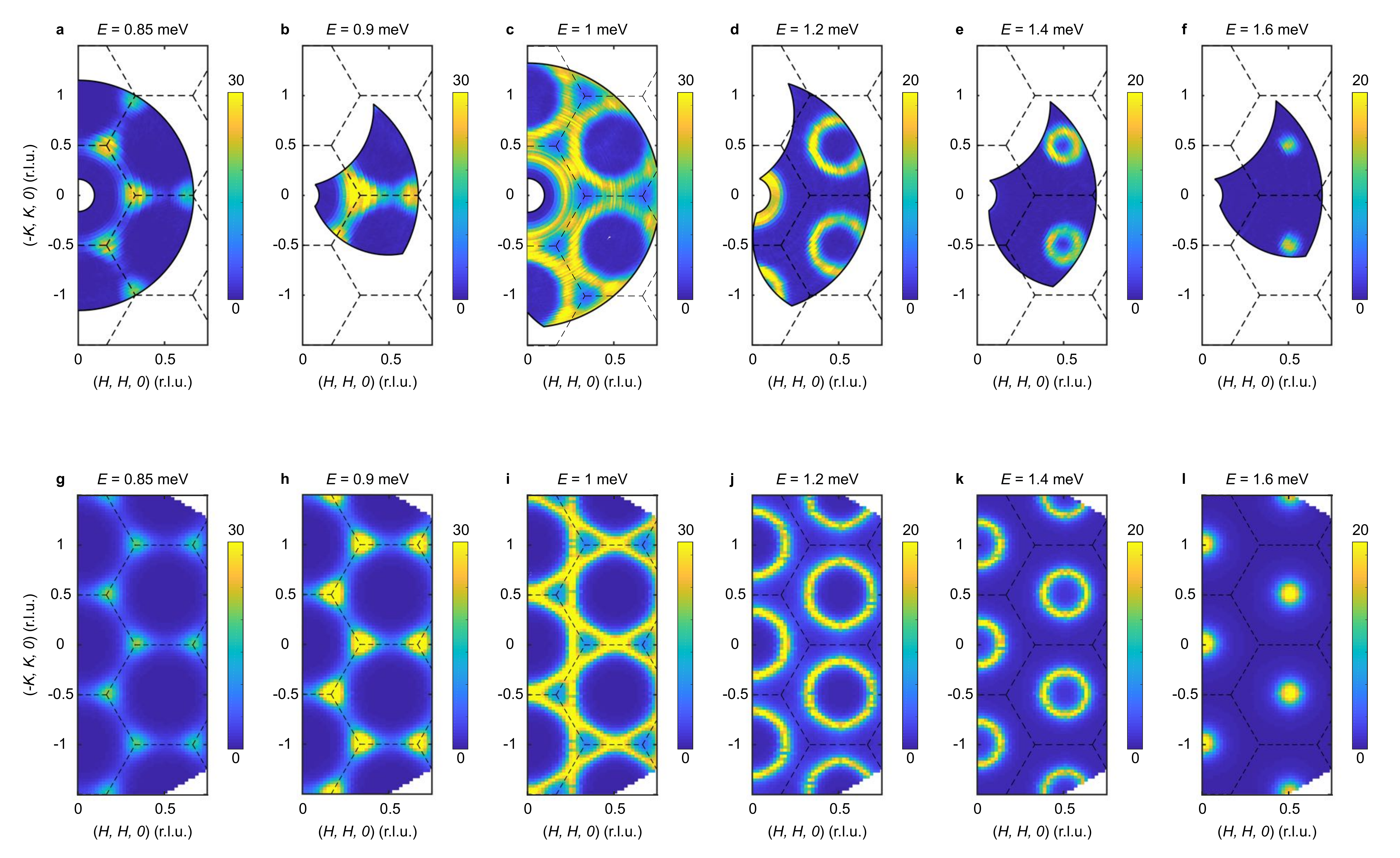}
\caption{
Measured and calculated momentum dependence of the CEF excitation in KTmSe$_2$
at the indicated energies and \emph{T} = 0.06 K. (a)-(f) Raw contour plots of the constant energy images
at \emph{T} = 0.06 K. (g)-(l) Calculated CEF excitation using the MF-RPA model specified in the text.
The dashed lines indicate the zone boundaries. The measurements were performed on ThALES triple-axis
spectrometer with $E_f$ = 4.066 meV.
}
\label{fig:2}
\end{figure*}

\section{Experimental method and results}
\label{sec2}

Polycrystalline samples of KTmSe$_2$ were synthesized through a solid-state reaction with a molar ratio of K : Tm : Se = 1 : 1 : 2. The starting materials were sealed in a vacuumed quartz tube and heated to 950$^\circ$C at a rate of 50$^\circ$C/h, then maintained at this temperature for three days. KTmSe$_2$ single crystals were grown by the flux method with a molar ratio of KTmSe$_2$ : K$_2$Se$_3$ = 1 : 10. The powder mixture was annealed at 950$^\circ$C for 5 hours, then slowly cooled down to 650$^\circ$C at a rate of 0.8$^\circ$C/h. The flux was subsequently removed with deionized water. KLuSe$_2$ single crystals, used as a non-magnetic reference, were grown using the same method. The resulting dark yellow KTmSe$_2$ single crystal exhibited a hexagonal shape with well-defined edges [Fig.~\ref{fig:1}(b)]. Single crystal X-ray diffraction measurements were performed on Bruker D8 X-ray diffractometer [Fig.~\ref{fig:1}(c)]. The X-ray powder diffraction data were refined using the FULLPROF SUITE software~\cite{Rodriguez1993}, from which the crystallographic parameters were extracted. A summary of the relevant crystallographic parameters for the KTmSe$_2$ is provided in Table~\ref{tab:latticeparamaters}.

Heat capacity, susceptibility, and magnetization measurements on KTmSe$_2$ single crystals were performed on a PPMS DynaCool instrument (Quantum Design). No signal of magnetic phase transition is observed in the heat capacity data under zero/non-zero external field along the crystalline \emph{c} axis, indicating the absence of long-range magnetic order in KTmSe$_2$ compound down to at least 0.3 K [Fig.~\ref{fig:1}(d)]. A broad peak appears at $\sim$ 4 K, and can be then partially suppressed by external magnetic fields, implying a large contribution from excited CEF states. The temperature dependent susceptibility shows Ising-like single-ion magnetic anisotropy with an evident easy axis and a negligible in-plane response. Moreover, the predominantly antiferromagnetic correlations are evidenced by the negative CW temperature of $\theta_c = - 2.32$ K (fit from 20 K to 50 K after subtracting the Van Vleck contribution) [Fig.~\ref{fig:1}(e)]. The isothermal magnetization data in both directions at 2 K reaffirms the Ising anisotropy, indicating that only the out-of-plane component requires examination [Fig.~\ref{fig:1}(f)]. Upon reaching 6 T, the magnetization saturates and increases linearly with the applied field. The saturated magnetic moment is $\mu_z$ = 6.2 $\mu_\textrm{B}$, from which we extract the \emph{g} factor $g_z$ = 1.033 ($\mu_z$ = $g_zJ\mu_\textrm{B}$ = 6$g_z\mu_\textrm{B}$). The \emph{g} factor is close to the Land\'{e} g-factor of Tm$^{3+}$ ion $g_J$ = 1.167, further confirming the Ising nature of KTmSe$_2$. Inelastic neutron scattering data were collected on 3 g of KTmSe$_2$ single crystals in the (\emph{H}, \emph{K}, 0) scattering plane. The measurements were carried out on the ThALES cold triple-axis spectrometer using flat-cone analyzer at the Institute Laue-Langevin, Grenoble, France.
\begin{table}[b]
\caption{\label{tab:latticeparamaters}
Refined crystallographic parameters from room-temperature X-ray powder diffraction with pulverized single-crystal KTmSe$_2$ samples.
}
\setlength{\tabcolsep}{9.0mm}{}
\begin{tabular}{cc}
\hline
\hline
\multicolumn{2}{c}{KTmSe$_2$ single crystal} \\
\multicolumn{2}{c}{(Trigonal, space group $R$-$3m$, Z = 3)} \\
\hline
\multicolumn{2}{c}{Lattice paramaters} \\
\hline
\emph{a} ({\AA}) & 4.1352(1)  \\
\emph{c} ({\AA}) & 22.7487(5)  \\
\emph{V} ({\AA}$^3$) & 336.881(2)  \\
$\alpha$ ($^\circ$) & {90} \\
$\gamma$ ($^\circ$) & {120} \\
\hline
\multicolumn{2}{c}{Atomic positions} \\
\hline
\multicolumn{2}{c}{K} \\
(\emph{x}, \emph{y}, \emph{z}) & {(0, 0, 0.5)} \\
Wyckoff site & b \\
Multiplicity & 3 \\

\multicolumn{2}{c}{Tm} \\
(\emph{x}, \emph{y}, \emph{z}) & {(0, 0, 0)} \\
Wyckoff site & a \\
Multiplicity & 3 \\

\multicolumn{2}{c}{Se} \\
(\emph{x}, \emph{y}, \emph{z}) & {[0, 0, 0.2698(1)]} \\
Wyckoff site & c \\
Multiplicity & 6 \\
\hline
\hline
\end{tabular}
\end{table}

\begin{figure}[b]
\includegraphics[width=0.49\textwidth]{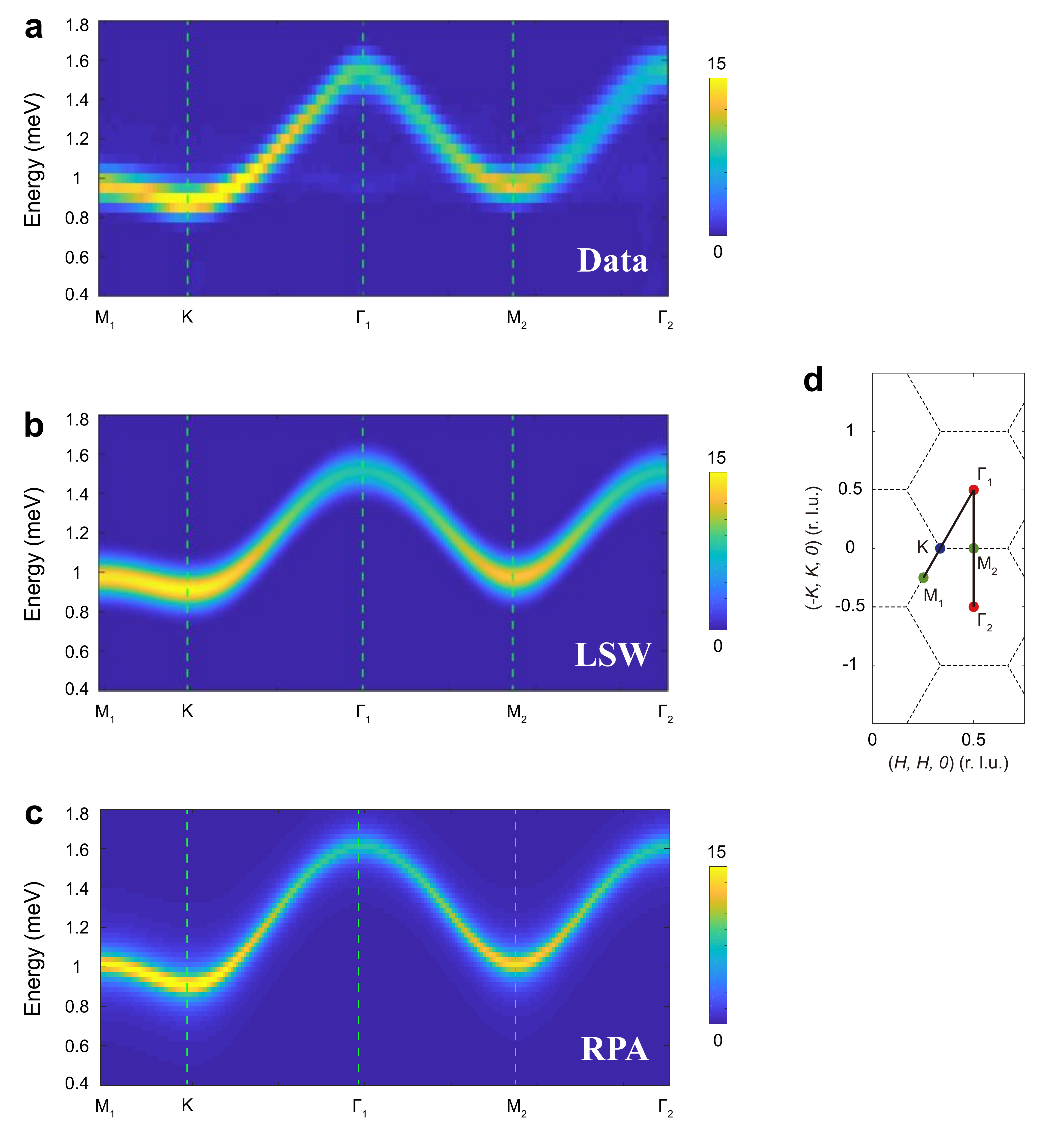}
\caption{
(a) Energy dependence of the observed spectral intensity along the high-symmetry momentum direction at \emph{T} = 0.06 K.
(b),(c) Simulated magnetic excitation spectra based on LSW and MF-RPA methods, respectively. (d) Schematic of a high-symmetry direction of KTmSe$_2$.
}
\label{fig:3}
\end{figure}

\begin{figure}[b]
\includegraphics[width=0.48\textwidth]{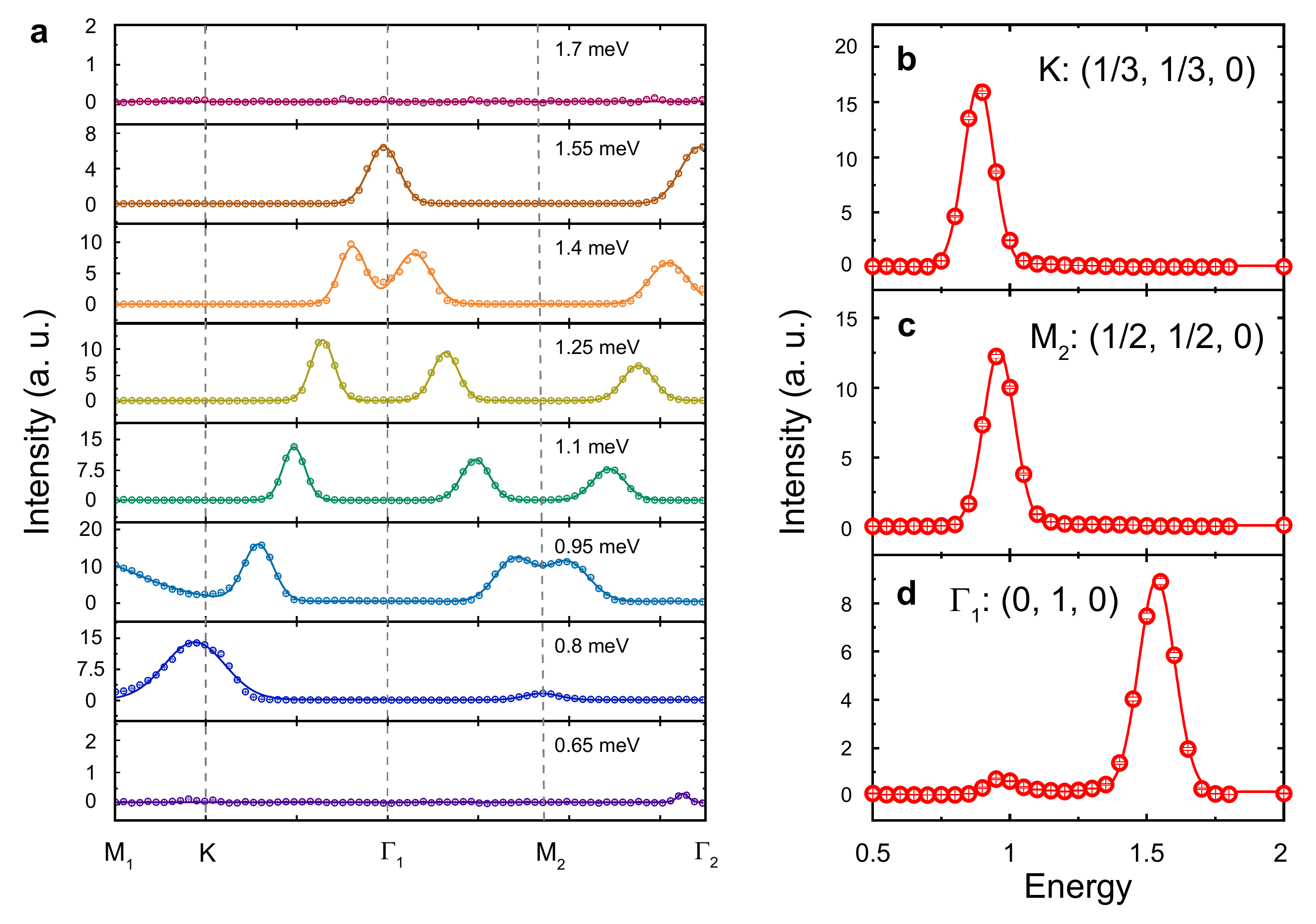}
\caption{
(Color online) (a) Constant energy cuts along the $M_1$-$K$-$\Gamma_1$-$M_2$-$\Gamma_2$ direction at indicated energies at \emph{T} = 0.06 K. The vertical dashed lines indicate the high-symmetry points. (b)-(d) Constant \emph{Q} cuts at $K$, $M_2$ and $\Gamma_1$ points, the error bars denote the standard deviation.
}
\label{fig:4}
\end{figure}

No magnetic Bragg peak is observed at high-symmetry points down to 60 mK in neutron diffraction measurements, confirming the absence of long-range magnetic order in KTmSe$_2$, as indicated by the heat capacity data.
With the increase of energy, a branch of dispersive magnetic excitation becomes visible at the
\emph{K} points above the energy gap of $\sim$ 0.85 meV [Fig.~\ref{fig:2}(a)].
The excitation disperses outward from the \emph{K} point [Fig.~\ref{fig:2}(b)] and then forms ring-like patterns
around the $\Gamma$ point at higher energies [Fig.~\ref{fig:2}(c)-(e)].
Eventually, the spectra reach the band top at the $\Gamma$ point and vanish above 1.6 meV
[Fig.~\ref{fig:2}(f)]. The overall dispersion can be seen more clearly in the energy dependence of the spectral intensity
along the high-symmetry directions [Fig.~\ref{fig:3}(a)]. The dispersion can be further determined quantitatively by constant-energy cuts [Fig.~\ref{fig:4}].

\begin{table*}
\caption{\label{tab:CEFparameters}
CEF parameters $B_n^m$ obtained from PC analysis. The units are in meV.
}
\begin{ruledtabular}
\begin{tabular}{cccccc}
$B^0_2$ & $B^0_4$ & $B^3_4$ & $B^0_6$ & $B^3_6$ & $B^6_6$ \\
\colrule
--1.54 $\times$ 10$^{-2}$ & --7.41 $\times$ 10$^{-4}$ & 2.12 $\times$ 10$^{-2}$ & --1.53 $\times$ 10$^{-6}$ & --2.42 $\times$ 10$^{-7}$ & --1.58 $\times$ 10$^{-5}$ \\
\end{tabular}
\end{ruledtabular}
\end{table*}

\section{Quantum Ising model description with lower doublets}
\label{sec3}

An effective spin-1/2 model is widely used to treat the low-energy magnetic excitations in many rare-earth magnets,
assuming that the magnetic properties at low temperatures are governed by the interaction between the $S_{\text{eff}}$ = 1/2
local moments with no involvement of higher excited CEF states. In the triangular-lattice compound TmMgGaO$_4$,
a quasi-doublet consisting of two lowest singlets
with an energy gap of $\sim$ 0.62 meV is taken considered as the magnetic ground state,
corresponding to a pseudospin-1/2 with an effective transverse field
along the \emph{y} direction~\cite{Shen2019,Liu2020}. In this picture,
the out-of-plane spin operator $\hat{S}^{z}$ represents the three-sublattice dipole order
while the in-plane operators $\hat{S}^{x}$ and $\hat{S}^{y}$ represent the ferroquadrupole order,
described with transverse field Ising model~\cite{Liu2020,Chen2019}.
We here apply the same quantum Ising model description to
explain the dispersive magnetic excitation in KTmSe$_2$ observed in INS.
Unlike TmMgGaO$_4$ where the exchange interaction wins and supports
an ordered state via the assistance of the intrinsic transverse Zeeman coupling
by an order-by-disorder mechanism~\cite{Shen2019,Liu2020},
however, the intrinsic transverse Zeeman coupling
wins and preserves the $\mathbb{Z}_2$ symmetry of the system in KTmSe$_2$.

In the magnetization measurements on single-crystalline samples, the
Tm$^{3+}$ ions are fully polarized under large magnetic fields (upon 8 T),
from which we find that the \emph{g} factor $g_z$ = 1.033 is close to the land\'{e} factor $g_J$ = 1.167.
The single-ion Ising anisotropy suggests that the ground state wavefunction
of KTmSe$_2$ is dominated by $|J^z = \pm6\rangle$,
which means the CEF ground state is a singlet~\cite{LiYS2020},
\begin{eqnarray}
|\Psi_g\rangle = c_6(|6\rangle + |{-6}\rangle) + c_3(|3\rangle
- |{-3}\rangle) + c_0|0\rangle
\label{eq:TFIMGSwavefunction},
\end{eqnarray}
additionally, in our heat capacity measurements, the full magnetic entropy release shows $\sim$ $R$ln2 around 15 K, indicating that the first excited CEF state is supposed to be another singlet,
\begin{eqnarray}
|\Psi_e\rangle = c'_6(|6\rangle - |{-6}\rangle) + c'_3(|3\rangle + |{-3}\rangle)
\label{eq:TFIM1stwavefunction}.
\end{eqnarray}

The CEF information obtained from characterizations suggests that the ground state of KTmSe$_2$ is expected to be a 'quasi-doublet', consisting of the CEF ground state singlet and the first excited singlet~\cite{Shen2019}, similar to the case of TmMgGaO$_4$. As a result, we can construct effective spin-1/2 operators acting on this quasi-doublet.
\begin{eqnarray}
&&\hat{S}^{x} = \frac{i}{2}(|\Psi_e\rangle\langle\Psi_g| - |\Psi_g\rangle\langle\Psi_e|)
\label{eq:spin operater_x}, \\
&&\hat{S}^{y} = \frac{1}{2}(|\Psi_g\rangle\langle\Psi_g| - |\Psi_e\rangle\langle\Psi_e|)
\label{eq:spin operater_y},\\
&&\hat{S}^{z} = \frac{1}{2}(|\Psi_g\rangle\langle\Psi_e| + |\Psi_e\rangle\langle\Psi_g|)
\label{eq:spin operater_z},
\end{eqnarray}
here $|\Psi_g\rangle$ and $|\Psi_e\rangle$ are eigenstates corresponded to eigenvalues of 1/2 and $-1/2$ of $\hat{S}^{y}$.
It has to be emphasized that the definition of the spin operators Eq.~\eqref{eq:spin operater_x} to ~\eqref{eq:spin operater_z} is to some extent different from the conventional meaning. The constructed longitudinal component $\hat{S}^{z}$ behaves as a magnetic dipole which couples to neutron or external fields, while the transverse components $\hat{S}^{x}$ and $\hat{S}^{y}$ are even under time-reversal and represent the multipolar behaviors that display `hidden' in conventional experimental probes~\cite{Shen2019,Liu2020}.

Based on the spin operators of the quasi-doublet,
we further establish the spin Hamiltonian of KTmSe$_2$
and adopt the linear spin wave (LSW) theory to calculate the spin excitation spectra,
\begin{eqnarray}
\hat{\mathcal{H}} = \sum_{\langle i,j\rangle}J_1\hat S_i^z\hat S_j^z
+ \sum_{\langle\langle i,j\rangle\rangle}J_2\hat S_i^z\hat S_j^z - h\sum_{i} \hat S_i^y
\label{eq:TFIMspinHAMILTONIAN},
\end{eqnarray}
where $\langle i,j\rangle$ and $\langle\langle i,j\rangle\rangle$ denote the nearest and next-nearest neighbors, respectively, and $h$ is the energy corresponding to the effective intrinsic transverse field describing the splitting of the ground state quasi-doublet,
as described with TFIM~\cite{Liu2018,Shen2019,Qin2022,Chen2019,Liu2020}.
A set of parameters can describe the neutron scattering data accurately: $J_1$ = 0.29(1) meV, $J_2$ = 0.018(2) meV, $h$ = 1.13(2) meV,
the calculated results reasonably agree with the experimental observations [Fig.~\ref{fig:3}(b)].

The successful application of the quantum Ising spin model both in TmMgGaO$_4$~\cite{Shen2019,Liu2020}
and KTmSe$_2$ provides a unique method to understand the magnetic excitation in Ising antiferromagnets
by introducing an effective transverse field to describe the splitting of ground state quasi-doublet.
Comparing these two materials, we found that TmMgGaO$_4$ is located in the three-sublattice ordered state
while KTmSe$_2$ is in the quantum disordered state that is driven by the dominant transverse exchange.
These results are consistent with the general phase diagram tackled with the Weiss mean-field approximation~\cite{Liu2019,Liu2020}.
Quite remarkably, the simple TFIM provides us an unified understanding on these two compounds:
since the quantum order-by-disorder effect in TmMgGaO$_4$ is relatively weak,
a coexistence of three-sublattice magnetic order and ``ferroquadrupole hidden order'' displays as a
nearly gapless magnetic excitation in INS, while a relatively large effective transverse field $h$
in KTmSe$_2$ results to the dominance of the polarization effect and suppression of
the three-sublattice order, then drives the system into the `quantum disordered' state
where the effective spins are fully polarized along the transverse direction.

\section{Full crystal field scheme calculation}
\label{sec4}

In the phase diagram of the TFIM, KTmSe$_2$ is located in the fully polarized region
where the three-sublattice ordering is fully suppressed by the intrinsic polarization effect.
This is consistent with the results from the heat capacity and neutron diffraction measurements.
Due to the absence of the long-range spin order, the dispersive excitation observed in the
INS can be also considered as the first excited CEF state from another perspective.
On the other hand, the spin exchange interaction between Tm$^{3+}$ ions still
play an important role and is responsible for the dispersion for CEF excitations.

\begin{table*}
\caption{\label{tab:PCeigen}
Calculated eigenvalues and eigenvectors from PC analysis ($\leq$5.45 {\AA}) on KTmSe$_2$.
}
\setlength{\tabcolsep}{9mm}{}
\begin{tabular}{l l}
\hline
\hline
Eigenvalue (meV) & Eigenvector \\
\hline
0 & 0.51($|6\rangle$ + $|-6\rangle$) + 0.354($|3\rangle$ -- $|-3\rangle$) + 0.479$|0\rangle$ \\
1.171 & --0.643($|6\rangle$ -- $|-6\rangle$) -- 0.294($|3\rangle$ + $|-3\rangle$) \\
2.816 & 0.235($|5\rangle$ + $|-5\rangle$) + 0.264($|4\rangle$ -- $|-4\rangle$) + 0.342($|2\rangle$ -- $|-2\rangle$) + 0.509($|1\rangle$ + $|-1\rangle$) \\
2.816 & 0.235($|5\rangle$ -- $|-5\rangle$) -- 0.264($|4\rangle$ + $|-4\rangle$) + 0.342($|2\rangle$ + $|-2\rangle$) -- 0.509($|1\rangle$ - $|-1\rangle$) \\
4.798 & 0.451($|6\rangle$ + $|-6\rangle$) -- 0.149($|3\rangle$ -- $|-3\rangle$) -- 0.74$|0\rangle$ \\
4.837 & 0.351($|5\rangle$ + $|-5\rangle$) -- 0.253($|4\rangle$ -- $|-4\rangle$) + 0.449($|2\rangle$ -- $|-2\rangle$) -- 0.333($|1\rangle$ + $|-1\rangle$) \\
4.837 & --0.351($|5\rangle$ -- $|-5\rangle$) -- 0.253($|4\rangle$ + $|-4\rangle$) -- 0.449($|2\rangle$ + $|-2\rangle$) -- 0.333($|1\rangle$ -- $|-1\rangle$) \\
12.394 & 0.294($|6\rangle$ -- $|-6\rangle$) -- 0.643($|3\rangle$ + $|-3\rangle$) \\
15.659 & 0.434($|5\rangle$ -- $|-5\rangle$) -- 0.424($|4\rangle$ + $|-4\rangle$) -- 0.275($|2\rangle$ + $|-2\rangle$) + 0.236($|1\rangle$ -- $|-1\rangle$) \\
15.659 & -0.434($|5\rangle$ + $|-5\rangle$) -- 0.424($|4\rangle$ -- $|-4\rangle$) + 0.275($|2\rangle$ -- $|-2\rangle$) + 0.236($|1\rangle$ + $|-1\rangle$) \\
15.888 & 0.191($|6\rangle$ + $|-6\rangle$) -- 0.594($|3\rangle$ -- $|-3\rangle$) + 0.472$|0\rangle$ \\
16.71 & -0.365($|5\rangle$ + $|-5\rangle$) + 0.432($|4\rangle$ -- $|-4\rangle$) + 0.325($|2\rangle$ -- $|-2\rangle$) -- 0.274($|1\rangle$ + $|-1\rangle$) \\
16.71 & 0.365($|5\rangle$ -- $|-5\rangle$) + 0.432($|4\rangle$ + $|-4\rangle$) -- 0.325($|2\rangle$ + $|-2\rangle$) -- 0.274($|1\rangle$ -- $|-1\rangle$) \\
\hline
\hline
\end{tabular}
\end{table*}

By establishing a CEF Hamiltonian below, we can then describe the CEF states of Tm$^{3+}$ quantitatively,
\begin{eqnarray}
\hat{\mathcal{H}}_\textrm{CEF} = \sum_{n,m} B^m_n\hat O^m_n
\label{eq:H-CEF},
\end{eqnarray}
where $\hat O_n^m$ are the Stevens operators and $B_n^m$ are multiplicative factors called the CEF parameters~\cite{Stevens1952}.
Under the D$_{3d}$ point symmetry, only six CEF parameters are allowed to be non-zero in [Eq.~(\ref{eq:H-CEF})]:
$B_2^0$, $B_4^0$, $B_4^3$, $B_6^0$, $B_6^3$, and $B_6^6$. The thirteen-fold ($J$ = 6, $L$ = 5, $S$ = 1)
degenerate ground state of the free 4$f^{13}$ Tm$^{3+}$ is split into five singlets and four doublets
under the $D_{3d}$ CEF environment with a rotation axis about $c$~\cite{LiYS2020}.
We have used the McPhase software package to carry out the PC analysis
based on the following formula~\cite{Rotter2004},
\begin{eqnarray}
B_n^m = \frac{4\pi}{2n+1}\frac{|e|^2}{4\pi\epsilon_0}\sum_{i}\frac{q_i}{r_i^{n+1}}a_0^n\langle{r^n}\rangle{Z_n^m}(\theta_i,\phi_i)
\label{eq:CEFpara_simulation}.
\end{eqnarray}

In our PC analysis, we selected a point-charge shell ($\leq$5.45 {\AA})
to describe the CEF environment of Tm$^{3+}$, including six nearest (2.79 {\AA})
and six next-nearest (4.99 {\AA}) Se$^{2-}$ anions, six nearest Tm$^{3+}$ (4.14 {\AA})
and K$^{+}$ (4.48 {\AA}) ions. Although the Se$^{2-}$ ionic shell in the TmSe$_6$ octahedra is responsible for the primary splitting of the Tm$^{3+}$ manifold, the resulting CEF states are often inaccurate when only six Se$^{2-}$ anions are considered in the CEF environment. This is because the neighboring Tm$^{3+}$ and K$^{+}$ ions also contribute significantly to the splitting~\cite{Bordelon2020}.  The simulated CEF parameters from the selected CEF
environment are listed in Table~\ref{tab:CEFparameters}, while the corresponding eigenvalues
and eigenvectors are listed in Table~\ref{tab:PCeigen}.

From the PC analysis, we again verified the CEF ground state and the first excited state
both to be singlets, dominated by $|{J^z = \pm6 }\rangle$,
\begin{align}
&|\Psi_0\rangle = 0.51(|6\rangle + |{-6}\rangle)
+ 0.35(|3\rangle - |{-3}\rangle) + 0.48|0\rangle,
\label{eq:GSwavefunction}\\
&|\Psi_1\rangle = - 0.64(|6\rangle - |{-6}\rangle)
- 0.29(|3\rangle + |{-3}\rangle),
\label{eq:1stwavefunction}
\end{align}
while the energy of $|\Psi_1\rangle$ is $\sim$ 1.17 meV.
The results agree with our thermodynamic and neutron scattering measurements well.

In TmMgGaO$_4$, the influence of higher CEF levels on low-temperature magnetism below 10 K is negligible due to the large energy gap of approximately 38.4 meV~\cite{LiYS2020}. However, in KTmSe$_2$, the three excited states (2.816, 4.798, 4.837 meV) are somewhat closer to the ground state quasi-doublet. This raises concerns about the appropriateness of restricting the analysis to the lower quasi-doublet. While this restriction may not be a serious issue for KTmSe$_2$, it could be problematic for other rare-earth magnets with weak crystal field splittings. To ensure a comprehensive analysis, we will focus on the effective magnetic Hamiltonian that includes all CEF levels~\cite{LiYS2015,Zhang2021-1},
\begin{center}
\begin{eqnarray}
\hat{\mathcal{H}}_\textrm{eff} &=& \hat{\mathcal{H}}_\textrm{CEF} + \hat{\mathcal{H}}_\textrm{spin}
\nonumber \\
&=& \sum_{n,m} B^m_n\hat O^m_n  + \sum_{\langle i,j \rangle}
J_{x} \hat J^x_i\hat J^x_j + J_{y} \hat J^y_i\hat J^y_j + J_{z} \hat J^z_i\hat J^z_j . \nonumber \\
\label{eq:H-eff}
\end{eqnarray}
\end{center}

Due to the apparent Ising anisotropy of Tm$^{3+}$, it is reasonable to ignore the in-plane spin-exchange terms,
then Eq.~(\ref{eq:H-eff}) can be written as
\begin{eqnarray}
\hat{\mathcal{H}}_\textrm{eff} = \sum_{n,m} B^m_n\hat O^m_n +
\sum_{\langle i,j\rangle}
J_z \hat J^z_i\hat J^z_j
\label{eq:H-eff-zz},
\end{eqnarray}
where $J_{z}$ is the nearest out-of-plane exchange parameter, $\hat J_i^z$ and $\hat J_j^z$ describe the original
$J = 6$ operators of the Tm$^{3+}$ moment.
By establishing the effective Hamiltonian, we then combine the contributions from all the excited CEF states and the spin-exchange interactions.
Based on $\hat{\mathcal H}_{\text{eff}}$, McPhase uses the mean field-random phase approximation to re-calculated the dispersive
excitation of the first excited CEF state and find the appropriate parameter that can accurately describe the present
neutron scattering data: $J_{z}$ = 0.004 meV. The calculated result of \emph{E}-\emph{k} dispersion shows an
excellent agreement with the experimental observation, as shown in Figs.~\ref{fig:2}(g)-(l), and~\ref{fig:3}(c).

The exchange parameter of 0.004 meV simulated from MF-RPA is much smaller than the that (0.29 meV)
obtained from TFIM, primarily because we focus on the effective total angular momentum ${J}$ = 6 rather than an effective spin ${S}_{\text{eff}}$ = 1/2.

The successful application of PC analysis and MF-RPA calculations to the effective Hamiltonian demonstrates the effectiveness of this method for analyzing CEF excitations in KTmSe$_2$. While TFIM can adeptly explain the magnetism by introducing an intrinsic transverse Zeeman term, MF-RPA offers an alternative, complementary approach to explain excitations.The methodology employed in this study, including the construction and simulation process along with appropriate anisotropic exchange parameters, could be adapted for use other rare-earth magnets. By providing a versatile approach to analyzing magnetic spectra, this method offers more opportunities to uncover the underlying nature of frustrated rare-earth magnets and explore novel phenomena that may emerge from their unique magnetic characteristics.

\section{Conclusion}
\label{sec5}

In conclusion, we have successfully synthesized the thulium-based triangular lattice magnet, KTmSe$_2$. Our comprehensive investigation, including thermodynamic and neutron scattering experiments, has suggested the absence of long-range magnetic order in this compound. Through magnetic susceptibility measurements, we have identified strong Ising-like interactions accompanied by antiferromagnetic correlations. Our inelastic neutron scattering measurements have unveiled a dispersive crystal field excitation branch, which can be accurately described by an effective spin-1/2 model with a transverse field. To further elucidate the CEF states and dispersive excitation, we have employed a point-charge analysis and mean field-random phase approximation based on the effective Hamiltonian, taking into account both the CEF interactions and spin exchange interactions. This study offers a comprehensive understanding of the low-temperature magnetism in KTmSe$_2$, as well as provides an effective approach for exploring the intricate magnetic behavior of materials containing 4$f$ electrons.


\begin{acknowledgments}
This work was supported by the Key Program of National Natural Science Foundation of China (Grant No. 12234006), National Key R\&D Program of China (2022YFA1403202), and the Shanghai Municipal Science and Technology Major Project (Grant No. 2019SHZDZX01). H. W. was supported by China National Postdoctoral Program for Innovative Talents (Grant No. BX2021080), China Postdoctoral Science Foundation (Grant No. 2021M700860), the Youth Foundation of the National Natural Science Foundation of China (Grant No. 12204108), and Shanghai Post-doctoral Excellence Program (Grant No. 2021481). G. C. was supported by the Research Grants Council of Hong Kong with General Research Fund (Grant No. 17306520).

\end{acknowledgments}

\bibliography{Reference} 

\end{document}